\documentstyle[aps,prl]{revtex}
\begin{document}
\draft
\twocolumn[\hsize\textwidth\columnwidth
\hsize\csname@twocolumnfalse\endcsname\draft\pagenumbering{roma}
\title{\Large\bf Adiabatic stabilization of excitons in
  intense terahertz laser}
\author{Ren-Bao Liu and Bang-Fen Zhu}
\address{Center for Advanced Study, Tsinghua University,
        Beijing 100 084, People's Republic of China}
\maketitle
\vspace{5mm} \begin{abstract}
High-precise calculation of near-infrared absorption
 and transient spectra in bulk semiconductors irradiated
by an intense terahertz (THz) laser shows that, the
ionization rate of the ground-state
exciton, probed through the dynamic Fano resonance,
may decrease with the increase of the THz laser intensity.
This counterintuitive effect indicates the excitons are
stabilized against the field-ionization. The much lower
Rydberg energy and ``atomic unit'' of laser intensity for
excitons, and the possibility of creating excitons from the
``vacuum state'' allow observing the excitonic stabilization
in experiments, in contrast to the case of atomic stabilization.
\end{abstract} \vspace{4mm}


\vskip 1pc ] \newpage \pagenumbering {arabic}

In accordance to Einstein's theory of the photoelectric effect,
an electron can be stripped from its atom by light beams with
sufficiently high frequency. As the intensity of a light beam
is increased, the stripping probability of the electron increases
owing to the increasing number of photon impacts. Surprisingly,
since the late 1980's, theoretical investigations and numerical
simulations\cite{XS1,XS2,XS3,XS4,XS5,science} have predicted that
when the electric field, associated with a laser of sufficiently
high intensity and frequency, approaches or exceeds the electrostatic
field between the electron and the ion in an atom, the wave function
of the irradiated atom may be distorted adiabatically into a
distribution with two well separated peaks. The peak spacing increases
with increasing the field intensity, thus the atomic electron
spends more time far away from the nucleus, and the  ionization
rate slows down dramatically till almost totally suppressed. In
other words, the adiabatic stabilization of the atom occurs.

The observation of this atomic stabilization effect is,  however, 
extremely difficult. Firstly, for the atomic ground state, the
high-frequency and high-intensity condition requires an extreme
ultraviolet laser with intensity larger than an atomic unit
($\sim 3.51\times 10^{16}$ W/cm$^2$), which seems not available
in the near future. Secondly, and more vitally, sufficiently slow
turn-on is demanded for the laser pulse to adiabatically drive the
atomic ground state into a stable dressed state; during the rise time
of the super-intense laser pulse, however, substantial ionization has
already been inevitable\cite{pulse}, which prevents the effect from
observation. Some researchers even argued that it was impossible to
observe such an adiabatic stabilization of atoms\cite{NOXS,NOXS1}.
In fact, except for a few disputable indications for the stabilization
of high Rydberg states\cite{experiment}, so far there is no unambiguous
experimental evidence for the atomic stabilization.

It is compelling to settle the debate whether this effect exists or
not. Apart from the pure curiosity in basic research, the atomic
stabilization is also of importance in potential application, such
as the high harmonic generation\cite{HHG} and quantum
computation\cite{Qbit}, as it can quench the undesired ionization
and dephasing caused by the intense laser that is used to drive
electrons or to manipulate a qubit in an atom.

In this Letter, we report on the adiabatic stabilization of
hydrogen-atom-like excitons in semiconductors. The excitonic
stabilization (XS), in contrast to its atomic counterpart, is
definitely an observable effect, owing to the basic characteristic
of excitons that they are created from ``vacuum'' of carriers by the
interband optical excitation. In our study, the dressed exciton
states are prepared by weak near-infrared (NIR) laser pulses in
semiconductors in the presence of a quasi-continuous terahertz (THz)
field, as in most recent experiments on the dynamical Franz-Keldysh
effect\cite{THz,SB,DFKE}. The ionization rate or lifetime of the
dressed excitons can be directly extracted by monitoring either the
linewidth of absorption peaks or the decay time of transient signals.
Thus the ``turn-on'' problem is safely circumvented as long as
the pulsed interband excitation and the following dephasing process
of excitons are well covered by the flattop region of the THz laser
pulse.

Moreover, the present laboratory gear is already ready for observing
the XS effect. Due to the small effective mass $\mu$ and large
dielectric constant $\varepsilon$, an exciton has much smaller binding
energy $E_B$ than the hydrogen atom does, and the ``excitonic unit''
for laser intensity ($\propto\mu^4/\varepsilon^{5.5}$) in a typical
semiconductor like GaAs is about 10 orders of magnitude smaller than
the atomic unit. So, the ``high-intensity and high-frequency'' requirement
for the XS can readily be fulfilled by the free-electron lasers operating
with MW/cm$^2$ power and THz frequency\cite{FEL}.

Our investigation is based on the high-precise calculation of
interband optical spectra, which includes properly the Coulomb
interaction, the nonperturbative ac-field, and the contribution of
continuum states. To avoid unnecessary complexity, the excitonic state
is treated within the simple effective-mass approximation. Then the
motion equation for the relative motion of an electron-hole pair is
\begin{eqnarray}
i\hbar\partial_t \psi({\bf r},t)=(\hat{H}-i\gamma_2)\psi({\bf r},t)
+\chi(t) e^{-i\Omega_0 t}\delta({\bf r}), \label {HH1} \\
\hat{H}\equiv E_g-{\hbar^2\over 2\mu}{\bf \nabla}^2_{\bf r}
-{e^2\over 4\pi\varepsilon r} +eFz\cos(\omega t), \label{HH}
\end{eqnarray}
where $E_g$ is the band gap, $F$ is the field strength of the THz laser
associated with frequency $\omega$  and linearly polarized in the
${\bf z}$ direction, $\chi (t)$ denotes the NIR pulse excitation
centered at $\Omega_0$, and $\gamma_2$ is an interband dephasing-rate
due to phonon scattering. The THz laser is assumed in a continuous
wave (cw) form, because its microsecond duration\cite{FEL} is much
longer than the picosecond interband excitation and dephasing process.
This model system is mathematically equivalent to a hydrogen atom
in an intense laser except that the electron-hole pair is generated
by the NIR laser.

Without the Coulomb coupling, the eigensolution of the
time-dependent Hamiltonian ( Eq. (\protect\ref{HH}))  is
the Volkov state\cite{Volkov}, which, as the basis set,  reads
$$
 |{\bf k},m\rangle=e^{im\omega t -ia_Fk_z\cos(\omega t)
 +i{U_p\over 2\hbar\omega}\sin(2 \omega t)} |\tilde{\bf k}\rangle,
$$
where $m$  and {\bf k}, associated with quasienergy
$\varepsilon_{{\bf k}m}={\hbar^2k^2\over 2\mu}+U_p+m\hbar\omega$,
denote the sideband index and the wavevector, respectively.
Here $U_p\equiv {e^2F^2 \over 4\mu\omega^2}$ is the effective
ponderomotive potential, $a_F\equiv {eF\over \mu \omega^2}$
is the classical excursion amplitude of an electron in the
ac-electric field, and $|\tilde{\bf k}\rangle$ is the Bloch-state with
accelerating quasimomentum ${\bf k}-{eF \over \hbar\omega}\hat{\bf z}
\sin(\omega t)$. With the excitonic Rydberg unit adopted,
the matrix elements of the Coulomb potential read
$$
V_{{\bf k}+{\bf q} m+n,{\bf k}m}
=-\pi^{-2} q^{-2}i^nJ_n(a_Fq_z),
$$
where $J_n(x)$ denotes the $n$th-order Bessel function of the first kind.
The  intra-sideband ($n=0$) interaction is directly related to the
Kramers-Hennerberg (KH) potential, the time-average of the Coulomb potential
in the coordinate frame rested on the quivering electron\cite{KH}.
The KH potential has logarithmic singularity in the segment
between the turning points $\pm \hat{\bf z} a_F$ and $r^{-1/2}$ singularity
at the ends. Consequently, the ground state $|\Phi\rangle$ is stretched
along the segment of singularity and becomes dichotomous for 
super-intense field ($a_F\gg a_B$, the exciton Bohr radius)\cite{XS1}.

The spectrum of the system exhibits the sideband structure.
As illustrated in the lower inset in Fig. \protect\ref{fig1}, each
sideband consists of several discrete states and a continuum,
and the discrete states are embedded in the ionization continua
of lower sidebands. As predicted very recently
in an ac-driven biased semiconductor superlattice\cite{DFR},
the quantum interference between a discrete state of quasienergy
excitons and the degenerate continuum of a sideband can lead to
the dynamic Fano resonance (DFR), which manifests itself as broadened
asymmetric lineshape in absorption spectra and as intrinsic decay in
transient four-wave mixing signals. Thus, the ionization rate of the
quasi-bound Floquet-state excitons and the XS effect can be studied
via the DFR induced by the inter-sideband coupling ($n\neq 0$) in the
present system.

To calculate the time-dependent wavefunction, we numerically integrate
Eq. (\protect\ref{HH1}) in the cylindrical polar coordinates
by extending  the space-time difference method\cite{Glutsch}
to a time-periodic system, with the initial condition as
$\psi({\bf r},-\infty)=0$. The out-going waves are absorbed
at the boundary with the mask technique\cite{XS4}. The NIR absorption
spectrum $\alpha(\Omega)\propto\Im \big[P(\Omega)/\chi(\Omega)\big]$,
in which $P(t)\equiv\psi({\bf 0},t)$ gives directly the transient
interband polarization (i.e. coherence). In the calculation,
$\hbar\omega=10$ meV ($\sim 2.4$ THz), $\gamma_2=1$ meV, and
the NIR pulse is assumed to be of a Gaussian shape,
i.e. $\chi(t)=\chi_0\exp\left(-{t^2/2\tau^{2}}\right)$.
For cw absorption, $\tau$ is taken as $8/\omega$, which is large enough
to eliminate the sideband overlap.  The material parameters
take bulk GaAs as an example ($E_B=4$ meV and $a_B=10$ nm).

Figure \protect\ref{fig1} plots the NIR cw-absorption spectra of
the bulk GaAs driven by the THz-field with various strength.
The blue-shift of the peak associated with the $1s$ exciton state
results mainly from the effective ponderomotive energy $U_p$ and
additionally from the decrease of the binding energy ${\cal E}_B$. The
sideband structure of the Floquet-state exciton is visible
in the spectra, and particularly, for $F>24$ kV/cm, or the
intensity $I>2.3$ MW/cm$^2$\cite{Note}, the $-1\omega$ sideband
can be stronger than the $0\omega$ one that evolves from the
original $1s$ state. There is also significant absorption in the
band gap. Indeed, the dynamical Franz-Keldysh effect\cite{DFKE}
is properly reproduced in this high-precise calculation, and the
result can be well understood in the Floquet-state picture.

The exciton ionization rate is probed through
the broadening and distortion of absorption peaks
in the NIR-spectra. As displayed in the inset,
the asymmetric resonance peak is well fitted with the Fano
lineshape\cite{FR} characterized by the shape parameter $q$
and broadening constant $\gamma$
$$
  \alpha(\Omega)=\alpha_0+\alpha_{1}
  {(q\gamma+\Omega-E_g+{\cal E}_B)^2 \over
        \gamma^2+(\Omega-E_g+{\cal E}_B)^2},
$$
indicating the occurrence of the DFR.
By fitting the $0\omega$ peak, the ionization rate $\Gamma$ can be
extracted via $\gamma\approx\gamma_2+\Gamma/2$.
As shown in Fig. \protect\ref{fig2}, $\gamma$
(hence the ionization rate) increases with the laser intensity
until $F\approx 16$ kV/cm (correspondingly, $I\approx 1.0$ MW/cm$^2$
\cite{Note} and $a_F\approx 1.28a_B$). Beyond that critical field
strength, the ionization rate commences to decrease, and approaches
zero at $F\approx 28$ kV/cm, clearly demonstrating the XS effect.

$\Gamma$ can also be  obtained analytically in the high-frequency
approximation\cite{Pont}. When the wavefunction dichotomy is
negligible for medium $a_F$ (say $a_F<3a_B$), a compact result
for $\Gamma$ of the ground state\cite{Pont} can be derived as
$$
{\Gamma}\approx \sum_{n\omega>{\cal E}_B} {32\pi \over k_n^3}
\left|\int^{+a_F}_{-a_F}{\Phi({\bf z}) dz\over 2a_F}\right|^2
\int_{0}^{1}  J^2_{n}(a_Fk_n x)dx,
$$
where $k_n^2=n\omega-{\cal E}_B$.
When field is enhanced so that $a_Fk_n$ approaches or
exceeds 1, due to rapid oscillation of the Bessel function,
the integration over $x$ diminishes. Physically, this originates
from the destructive interference of the out-going waves scattered
from different part of the initial state.
On the other hand, the stretching of the state along the field will
reduce the wavefunction peak, contributing additionally to the XS.

Fano interference is one of fundamental channels for irreversible
decay\cite{TRFR1,TRFR2}, so the XS can also be
directly observed in the time domain by monitoring the transient
optical signal after a pulsed excitation. The time-resolved interband
polarization is calculated for several THz-field strengths and plotted
in Fig. \protect\ref{fig3}. The duration of the NIR pulse $\tau$ is
then set to be $2/\omega$, and the central frequency is chosen for
each field strength to be resonant with the brightest transition,
namely $\Omega_0-E_g=-4$, $-2$, 2, 5, and 6 meV for $F=0$, 8, 16, 20,
and 30 kV/cm, respectively. To single out the dephasing induced by
the DFR only, the reformulated signal
${\cal P}(t)\equiv P(t) \exp(\gamma_2 t)$ is plotted.
At finite field, the optical signal  roughly  oscillates
with the period of $T\equiv 2\pi/\omega$, because the states in the
KH potential oscillate periodically in the laboratory frame.
The modulation beat superimposed on the fast oscillation
results from the quantum interference between the
Floquet states with different quasienergy. The dynamic Fano interference
between the discrete state and the continuum opens a new channel for
interband dephasing, leading to the decay of the transient signal.
The XS effect is verified by the fact that the dephasing
rate calculated for $F>16$ kV/cm begins to decrease, consistent
with the variation trend of the ionization rate as 
shown in Fig. \protect\ref{fig2}. For $F=30$ kV/cm
($I\approx 3.6$ MW/cm$^2$\cite{Note}), the decay
of the signal almost disappear.
To measure the transient signals, nonlinear optical experiments
like four-wave mixing are usually adopted, in which
many-body correlation is important sometimes\cite{Chemla}  and may induce
new interesting effects on XS, which, however, is out of the scope of our
present study.

Fig. \protect\ref{fig4} displays  snapshots of the
probability distribution of the exciton wavepacket
resonantly excited by a pulse ($\Omega_0-E_g=8.88$ meV) with
$\tau=20/\omega$, corresponding to the spectral width 0.5 meV.
When the NIR-excitation pulse is over, the
wavepacket performs almost perfect periodic swinging along the
$z-$axis with the turning points located at about
$\pm\hat{\bf z}a_F$, except for an overall exponential decay due to the
ionization. This periodic behavior is verified by the time-dependence
of the optical signal (see the inset). Because of the inter-sideband
mixing, the wavepacket shape varies drastically within a period,
in contrast to the unchanged shape of the periodically oscillating
wavepacket in the KH potential without inter-sideband coupling\cite{XS1}.
All these demonstrate that a quasi-bound Floquet state is formed in the
strong THz-field, and justify  our ascribing the suppression of the
Fano broadening and dephasing rate to the adiabatic XS.

In summary,  the novel excitonic stabilization effect, a counterpart of
the atomic stabilization effect, has been predicted and explored
nonperturbatively in a coupled system of semiconductors and
intense THz-fields by high-precise calculation.

This work was supported by the National Science Foundation of China, and
R.B.L was also supported by China Postdoctoral Science Foundation.

\begin{figure}
\caption{Linear absorption spectrum for several THz-field strengths
indicated by $F$. The upper-right inset shows, as an example, the
Fano resonance fitting process for $F=24$ kV/cm, where the squares are
the fitted data. The lower-right inset shows schematically the
sideband structure of the Floquet-state excitonic spectrum.}
\label{fig1}
\end{figure}

\begin{figure}
\caption{Fano broadening constant versus the THz-field strength.}
\label{fig2}
\end{figure}

\begin{figure}
\caption{Real-time dependence of the interband polarization for various
THz-field strength indicated by $F$, in which the curves are offset for
clarity. The intensity profile of the NIR pulse (the dotted curve)
is also shown for comparison.}
\label{fig3}
\end{figure}

\begin{figure}
\caption{Probability distribution of the NIR-pulse-excited wavepacket
at several time instant within a period $T$ for the THz-field of
$F=24$ kV/cm. The inset shows the semi-logarithmic plots of the
intensity of the NIR pulse (dotted line) and transient interband
optical signal ($|{\cal P}(t)|^2$, solid line) as functions of time, and
the time snatch (marked by the bar) where the snapshots are made.}
\label{fig4}
\end{figure}


\begin{references}
\bibitem{Note} Here we use the intensity inside the material with relative
 dielectric constant $\varepsilon/\varepsilon_0=9.0$. The vacuum value is
 larger by about 1.3.
\end{references}
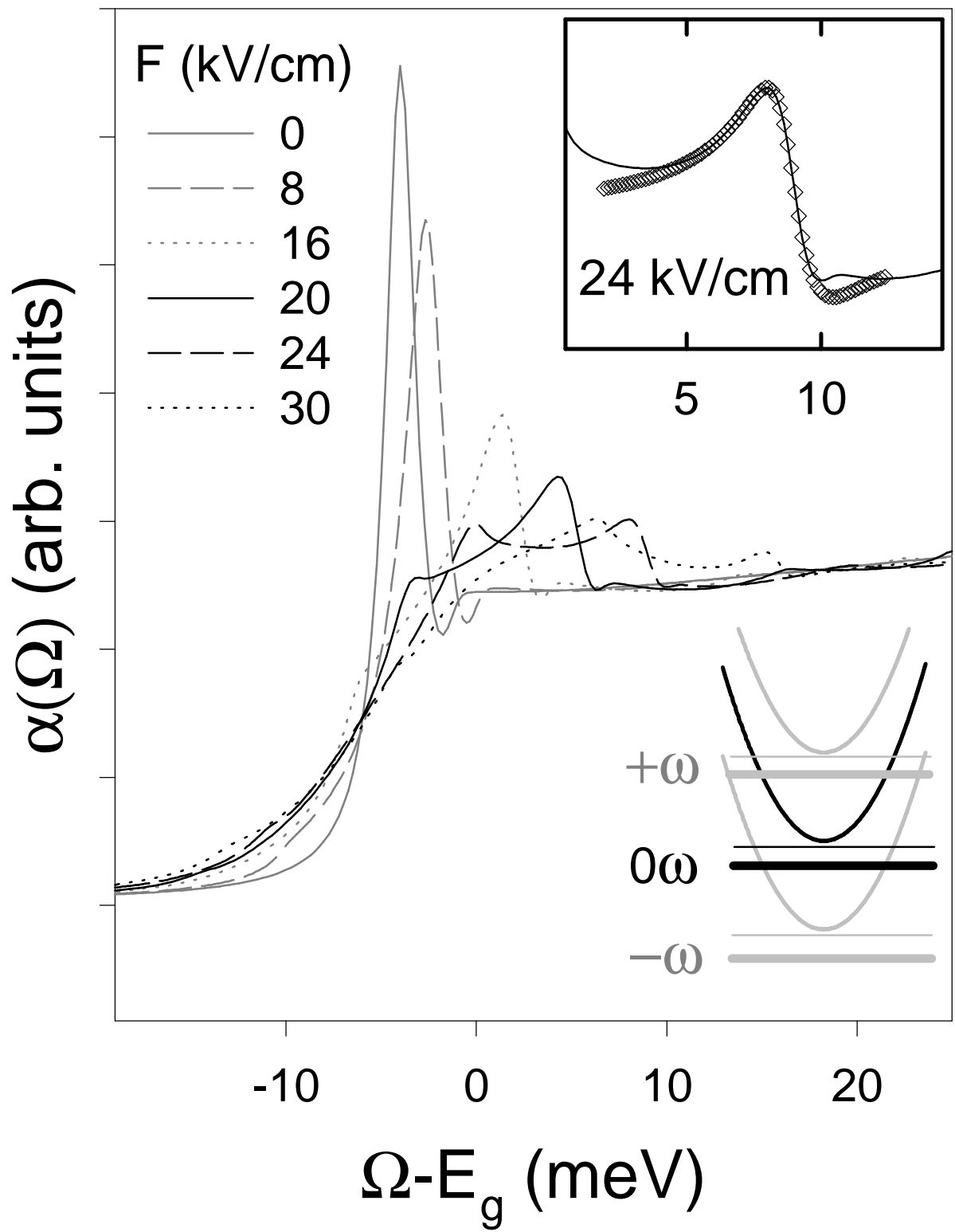
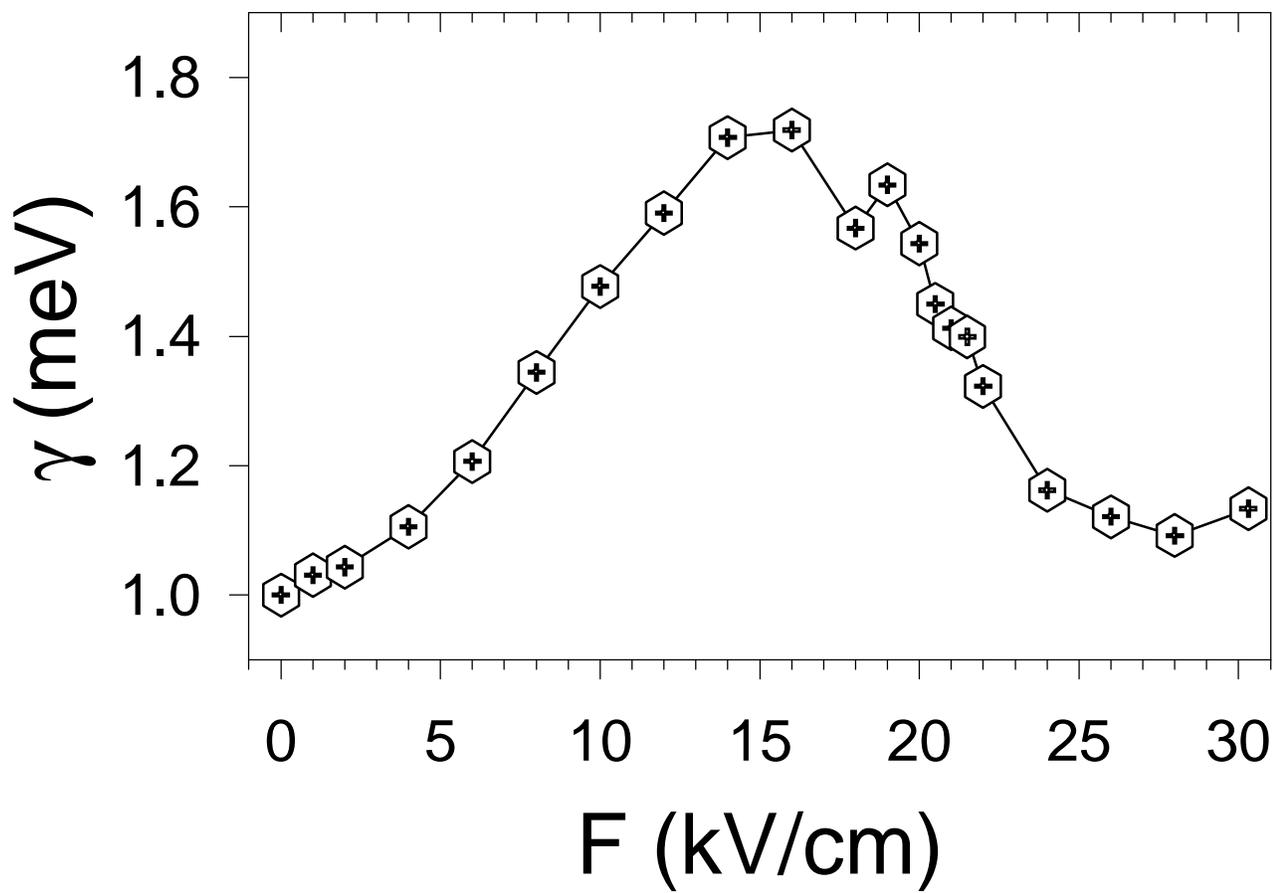
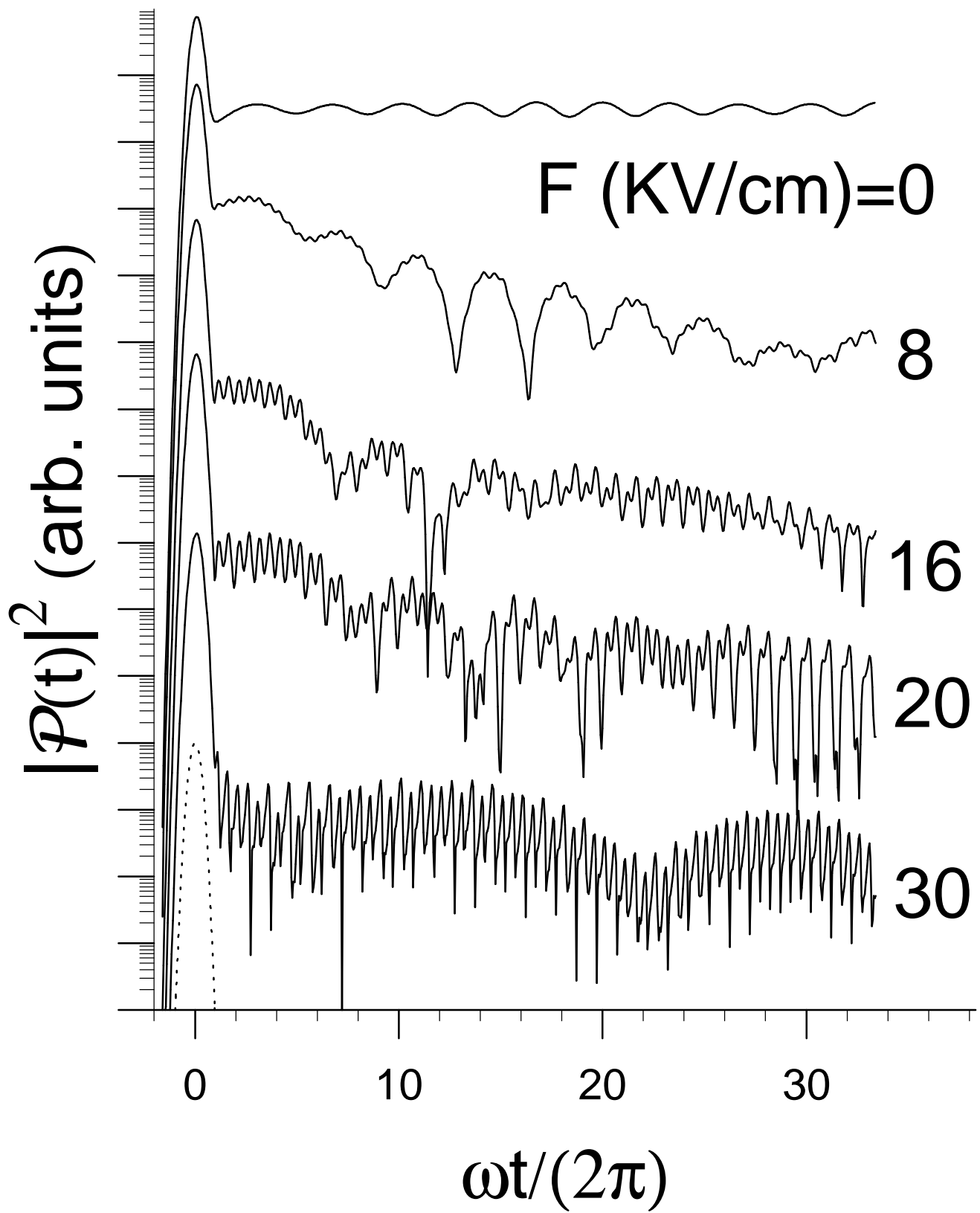
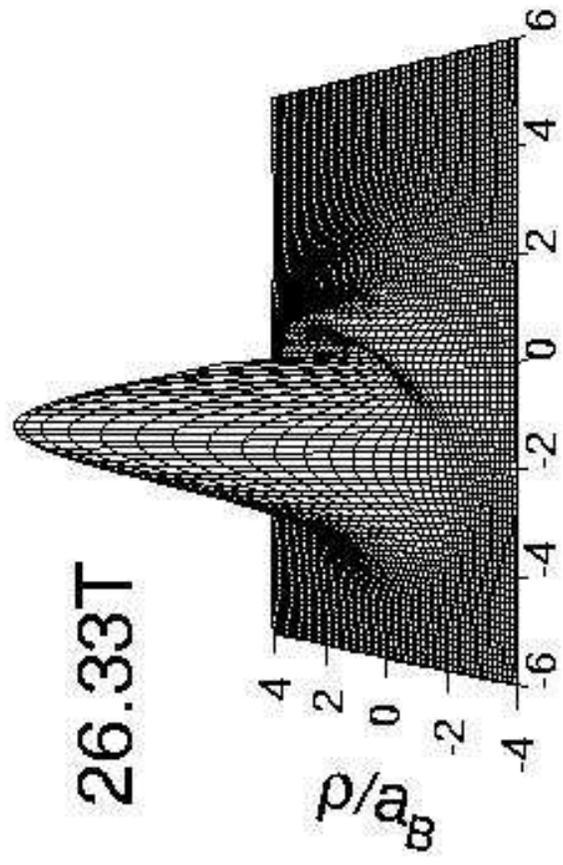
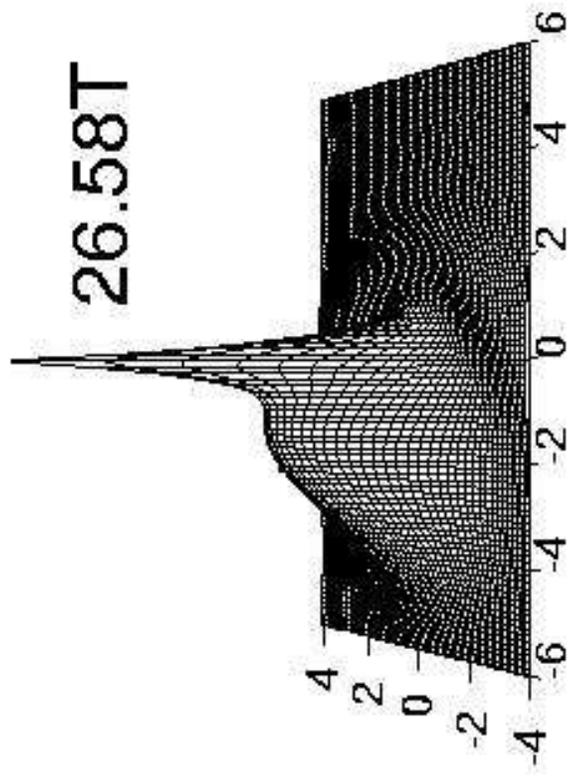
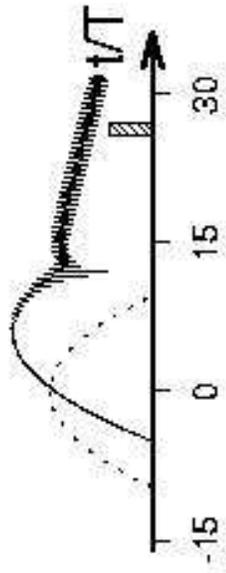
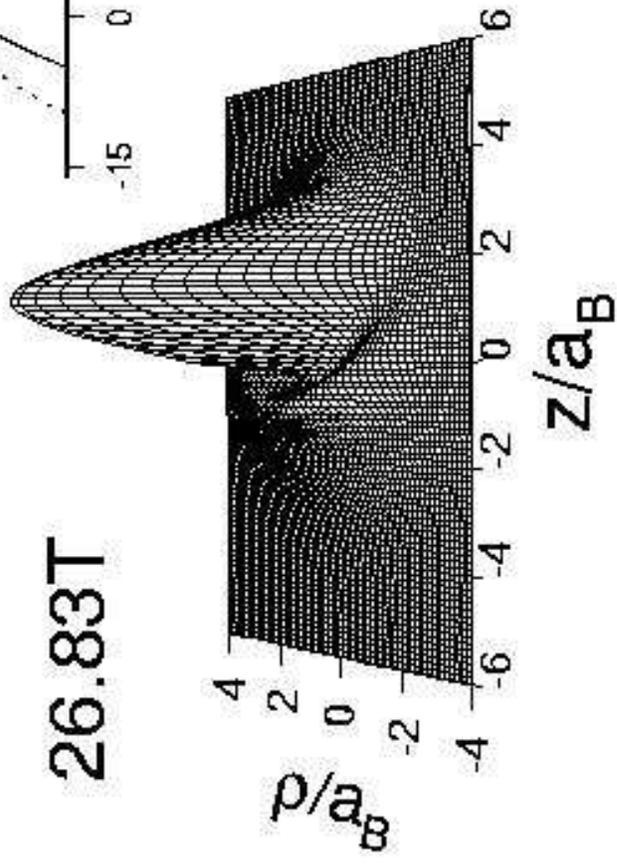
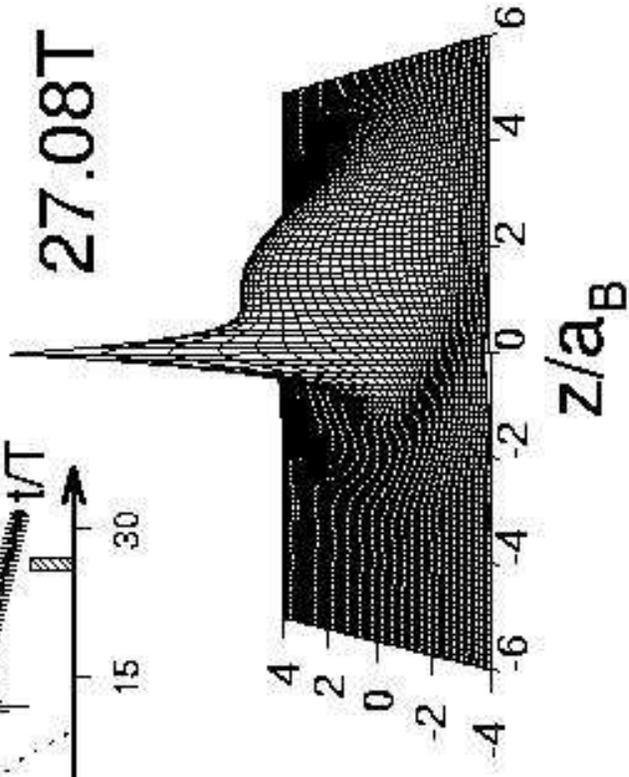
\end{document}